\documentclass[nofootinbib,aps,11pt]{revtex4-1}
\usepackage{graphicx}
\usepackage{hyperref}
\usepackage{cancel}
\usepackage{amssymb}
\usepackage{textcomp}
\usepackage{amsmath}
\usepackage{bm}
\usepackage{times}
\usepackage{epsfig}
\usepackage{color}

\begin{document}
\title{\LARGE Same-Sign Tetralepton Signature in Type-II Seesaw at Lepton Colliders}
\bigskip
\author{Xu-Hong Bai}
\author{Zhi-Long Han}
\email{sps\_hanzl@ujn.edu.cn}
\author{Yi Jin}
\author{Hong-Lei Li}
\author{Zhao-Xia Meng}
\email{sps\_mengzx@ujn.edu.cn}
\affiliation{
School of Physics and Technology, University of Jinan, Jinan, Shandong 250022, China}
\date{\today}

\begin{abstract}
	The same-sign tetralepton signature via mixing of neutral Higgs bosons and their cascade decays
	to charged Higgs bosons is a unique signal in the type-II seesaw model. In this paper, we study this signature at future lepton colliders, such as ILC,  CLIC, and MuC. Constrained by direct search, $H^{\pm\pm}\to W^\pm W^\pm$ is the only viable decay mode for $M_{A^0}=400$ GeV at $\sqrt{s}=1$ TeV ILC. With an integrated luminosity of $\mathcal{L}=8~ \mathrm{ab}^{-1}$, the promising region with about 150 signal events corresponds to a narrow band in the range of $10^{-4}~\text{GeV}\lesssim v_\Delta \lesssim10^{-2}$ GeV. For heavier triplet scalars $M_{A^0}\gtrsim 900$ GeV, although the $H^{\pm\pm}\to \ell^\pm \ell^\pm$ decay mode is allowed, the cascade decays are suppressed. A maximum event number $\sim 16$ can be obtained around  $v_\Delta\sim4\times10^{-4}$ GeV and $\lambda_4\sim0.26$ for $M_{A^0}=1000$ GeV with  $\mathcal{L}=5~ \mathrm{ab}^{-1}$ at $\sqrt{s}=3$ TeV CLIC. Meanwhile, we find that this signature is not promising for $M_{A^0}=1500$ GeV at $\sqrt{s}=6$ TeV MuC.

\end{abstract}

\maketitle

\section{Introduction}

The discovery of neutrino oscillations \cite{Fukuda:1998mi,Ahmad:2002jz,An:2012eh} confirms that neutrinos have sub-eV masses. Meanwhile, the underlying mechanism accounting for such tiny neutrino mass is still an open question. Regarding the standard model (SM) as a low energy effective field theory, the simplest pathway to generate neutrino mass is via the Weinberg operator $LL\Phi\Phi/\Lambda$ \cite{Weinberg:1979sa}. There are three possible ways at tree level to realize this operator \cite{Ma:1998dn}, which correspond to the canonical type-I \cite{Minkowski:1977sc,Mohapatra:1979ia}, type-II \cite{Magg:1980ut,Cheng:1980qt,Lazarides:1980nt,Mohapatra:1980yp}, and type-III  \cite{Foot:1988aq} seesaw. To verify whether these scenarios are realized in nature, the signatures of seesaw models at colliders have been extensively studied \cite{delAguila:2008cj,Deppisch:2015qwa,Cai:2017mow}. Since the conventional type-I seesaw requires the right-hand neutrinos $N$ to be quite heavy ($\gtrsim10^{14}$  GeV), it is far beyond the reach of current and future planed colliders. Therefore, we consider the type-II seesaw in this work. Other possible low scale approaches to generate tiny neutrino mass have been summarized in Ref.~\cite{Boucenna:2014zba,Cai:2017jrq}.

The type-II seesaw introduces a scalar triplet $\Delta$ with hypercharge $Y=+2$, where neutrino mass is generated by the Yukawa interaction between the lepton doublets and scalar triplet. After the spontaneous symmetry breaking of SM Higgs doublet $\Phi$, the trilinear term $\mu \Phi^T i\tau_{2}\Delta^\dagger\Phi$ induces a vacuum expectation value for the neutral component of scalar triplet with $v_\Delta\sim \mu v^2/M_\Delta^2$. Since the scalar triplet $\Delta$ also carries the lepton number $+2$, the $\mu$-term breaks the lepton number by two units. In particular, this trilinear term is the only source of lepton number violation, thus it should be naturally small. Then, for $\mu\sim v_\Delta$, we can naturally have $M_\Delta\sim v$, i.e., the mass of scalar triplet at the electroweak scale \cite{Arhrib:2011uy}.

A distinct feature of this model is the presence of doubly charged Higgs $H^{\pm\pm}$. 
Assuming degenerate mass spectrum of the scalar triplet, the typical channels to hunt for $H^{\pm\pm}$ are the same-sign dilepton channel $H^{\pm\pm}\to\ell^\pm\ell^\pm$ and the same-sign diboson channel $H^{\pm\pm}\to W^\pm W^\pm$ \cite{Perez:2008ha}. For non-degenerate case, cascade decay channel $H^{\pm\pm}  \to H^\pm W^\pm$ is also possible \cite{Melfo:2011nx,Aoki:2011pz,Han:2015hba,Han:2015sca}. Corresponding signatures have been extensively studied at LHC \cite{Akeroyd:2011zza,Akeroyd:2012nd,Chun:2013vma,Babu:2016rcr,Li:2018jns,Primulando:2019evb}, HE-LHC \cite{Du:2018eaw,deMelo:2019asm,Padhan:2019jlc,Fuks:2019clu}, $e^+e^-$ collider \cite{Blunier:2016peh,Agrawal:2018pci},  and $ep$ colliders \cite{Dev:2019hev,Yang:2021skb}. When $v_\Delta<10^{-4}$ GeV, the $H^{\pm\pm}\to \ell^\pm \ell^\pm$ is the dominant decay mode, and direct search at LHC has already excluded the region $M_{H^{\pm\pm}}<870$ GeV  \cite{Aaboud:2017qph}.  In this case, the branching ratios of $H^{\pm\pm}\to\ell^\pm\ell^\pm$ are only correlated with neutrino oscillation parameters \cite{Akeroyd:2007zv}. When $v_\Delta>10^{-4}$ GeV, the $H^{\pm\pm}\to W^\pm W^\pm$ mode becomes the dominant one, and searches for pair production of $H^{\pm\pm}$ in this diboson channel have excluded $M_{H^{\pm\pm}}<350$ GeV  \cite{Aaboud:2018qcu,Aad:2021lzu}. 

Among various possible collider signatures of the type-II seesaw, a unique one is the same-sign tetralepton signature  \cite{Chun:2012zu,Chun:2019hce}, which arises from the mixing of neutral Higgs bosons and their cascade decays to singly and doubly charged Higgs bosons. Previous studies \cite{Chun:2012zu,Chun:2019hce} focus on the hadron colliders as LHC and FCC-hh with $\sqrt{s}=100$ TeV. In this paper, we will analyze this signature at future lepton colliders. Considering current lower bound on doubly charged Higgs $M_{H^{\pm\pm}}>350$ GeV, this signature is beyond the reach of CEPC \cite{CEPCStudyGroup:2018ghi}. In order to pair produce $H^{\pm\pm}$, the collision energy should be at least higher than 700 GeV. Therefore, we take the following three benchmark scenarios to illustrate, i.e., $M_{H^{\pm\pm}}\sim 400$ GeV at $\sqrt{s}=1$ TeV ILC \cite{Barklow:2015tja,Fujii:2017vwa},  $M_{H^{\pm\pm}}\sim 1000$ GeV at  $\sqrt{s}=3$ TeV CLIC \cite{Linssen:2012hp,Robson:2018zje}, and $M_{H^{\pm\pm}}\sim 1500$ GeV at $\sqrt{s}=6$ TeV Muon Collider (MuC)  \cite{Delahaye:2019omf,Long:2020wfp}.

In our paper, the type-II seesaw model will be briefly introduced in Sec.~\ref{SEC2}. The branching ratios of the scalar triplet components are also discussed in Sec.~\ref{SEC2}. The same-sign tetralepton signals at ILC, CLIC, and MuC are analyzed in Sec.~\ref{SEC3}. Finally, the conclusion is presented in Sec.~\ref{SEC4}.

\section{The Model}\label{SEC2}
We concisely review the type-II seesaw in this section. Besides the SM Higgs doublet $\Phi$, a scalar triplet $\Delta$ is also employed, which can be denoted as
\begin{equation}
\Phi = \left(\begin{array}{c}
\phi^+\\ \Phi^0
\end{array}\right),\quad
\Delta=\left(\begin{array}{cc}
\frac{\Delta^{+}}{\sqrt{2}} & \Delta^{++} \\
\Delta^{0} & -\frac{\Delta^{+}}{\sqrt{2}}
\end{array}\right),
\end{equation}
where after spontaneous symmetry breaking, the neutral components can be further written as $\Phi^{0}=\frac{1}{\sqrt{2}}\left(v+\phi^{0}+i \chi^{0}\right)$ and $\Delta^{0}=\frac{1}{\sqrt{2}}\left(v_\Delta+\delta^{0}+i \eta^{0}\right)$, respectively.
The Yukawa interaction that generates tiny neutrino mass is given by
\begin{equation}
\mathcal{L}_{Y}=Y_{\Delta} \overline{L_{L}^{c}} i \tau_{2} \Delta L_{L}+\text { h.c. }
\end{equation}
The scalar potential involving $\Phi$ and $\Delta$ is
\begin{equation}
\begin{aligned}
V(\Phi, \Delta) &=m_{\Phi}^{2} \Phi^{\dagger} \Phi+M^{2} \operatorname{Tr}\left(\Delta^{\dagger} \Delta\right)+\left(\mu \Phi^{\mathrm{T}} i\tau_{2} \Delta^{\dagger} \Phi+\mathrm{h.c.}\right)+\frac{\lambda_0}{4}\left(\Phi^{\dagger} \Phi\right)^{2} \\
&+\lambda_{1}\left(\Phi^{\dagger} \Phi\right) \operatorname{Tr}\left(\Delta^{\dagger} \Delta\right)+\lambda_{2}\left[\operatorname{Tr}\left(\Delta^{\dagger} \Delta\right)\right]^{2}+\lambda_{3} \operatorname{Tr}\left[\left(\Delta^{\dagger} \Delta\right)^{2}\right]+\lambda_{4} \Phi^{\dagger} \Delta \Delta^{\dagger} \Phi.
\end{aligned}
\end{equation}
Mixing between the doublet and triplet scalars leads to seven physical scalars,i.e., doubly charged Higgs $H^{\pm\pm}$, singly charged Higgs $H^\pm$, CP-even Higgs bosons $h$ and $H^0$, and CP-odd Higgs $A^0$, with the mixing angles specified by
\begin{equation}
\tan \beta_{\pm}=\frac{\sqrt{2} v_{\Delta}}{v}, \quad \tan \beta_{0}=\frac{2 v_{\Delta}}{v}, \quad \tan 2 \alpha=\frac{4 v_{\Delta}}{v} \frac{v^{2}\left(\lambda_{1}+\lambda_{4}\right)-2 M_{\Delta}^{2}}{v^{2} \lambda-2 M_{\Delta}^{2}-4 v_{\Delta}^{2}\left(\lambda_{2}+\lambda_{3}\right)},
\end{equation}
where $M^2_\Delta=\mu v^2/(\sqrt{2}v_\Delta)$. The masses of the doubly and singly charged Higgs bosons $H^{++}$ and $H^{+}$ is given by
\begin{equation}
M_{H^{++}}^{2}=M_{\Delta}^{2}-v_{\Delta}^{2} \lambda_{3}-\frac{\lambda_{4}}{2} v^{2}, \quad M_{H^{+}}^{2}=\left(M_{\Delta}^{2}-\frac{\lambda_{4}}{4} v^{2}\right)\left(1+\frac{2 v_{\Delta}^{2}}{v^{2}}\right).
\end{equation}
The masses of CP-even Higgs bosons $h$, and $H^0$ can be written as
\begin{align}
M_{h}^{2}= \mathcal{T}_{11}^{2} \cos ^{2} \alpha+\mathcal{T}_{22}^{2} \sin ^{2} \alpha-\mathcal{T}_{12}^{2} \sin 2 \alpha, \\
M_{H^{0}}^{2}=\mathcal{T}_{11}^{2} \sin ^{2} \alpha+\mathcal{T}_{22}^{2} \cos ^{2} \alpha+\mathcal{T}_{12}^{2} \sin 2 \alpha,
\end{align}
where $\mathcal{T}_{11}$, $\mathcal{T}_{22}$, and $\mathcal{T}_{12}$ are of the form
\begin{equation}
\mathcal{T}_{11}^{2}=\frac{\lambda_0}{2} v^{2} , \quad \mathcal{T}_{22}^{2}=M_{\Delta}^{2}+2 v_{\Delta}^{2}\left(\lambda_{2}+\lambda_{3}\right), \quad \mathcal{T}_{12}^{2}=-\frac{2 v_{\Delta}}{v} M_{\Delta}^{2}+ v_{\Delta}v\left(\lambda_{1}+\lambda_{4}\right).
\end{equation}
Finally, the CP-odd Higgs $A^0$ has the following mass
\begin{equation}
M_{A^0}^{2}=M_{\Delta}^{2}\left(1+\frac{4 v_{\Delta}^{2}}{v^{2}}\right).
\end{equation}

Constrained by the $\rho$ parameter, $v_\Delta\lesssim 1$ GeV should be satisfied. Neglecting the contributions from $v_\Delta$, masses of triplet scalars have the relation
\begin{equation}
	M_{H^{++}}^2-M_{H^+}^2\approx M_{H^{+}}^2-M_{H^0,A^0}^2\approx -\frac{1}{4}\lambda_4v^2.
\end{equation}
In this paper, we consider the scenario with $\lambda_4>0$, which leads to the mass spectrum $M_{H^{++}}<M_{H^+}<M_{H^0}\simeq M_{A^0}$. The mass difference between $H^0$ and $A^0$ plays a vital important role in the production of the same-sign trilepton signature, which is controlled by $v_\Delta$ as 
\begin{equation}
	M_{H^0}^2-M_{A^0}^2\sim 2(\lambda_2+\lambda_3)v_\Delta^2-4\frac{M_\Delta^2}{v^2}v_\Delta^2.
\end{equation} 

\begin{figure} 
	\centering
	\includegraphics[width=1\textwidth]{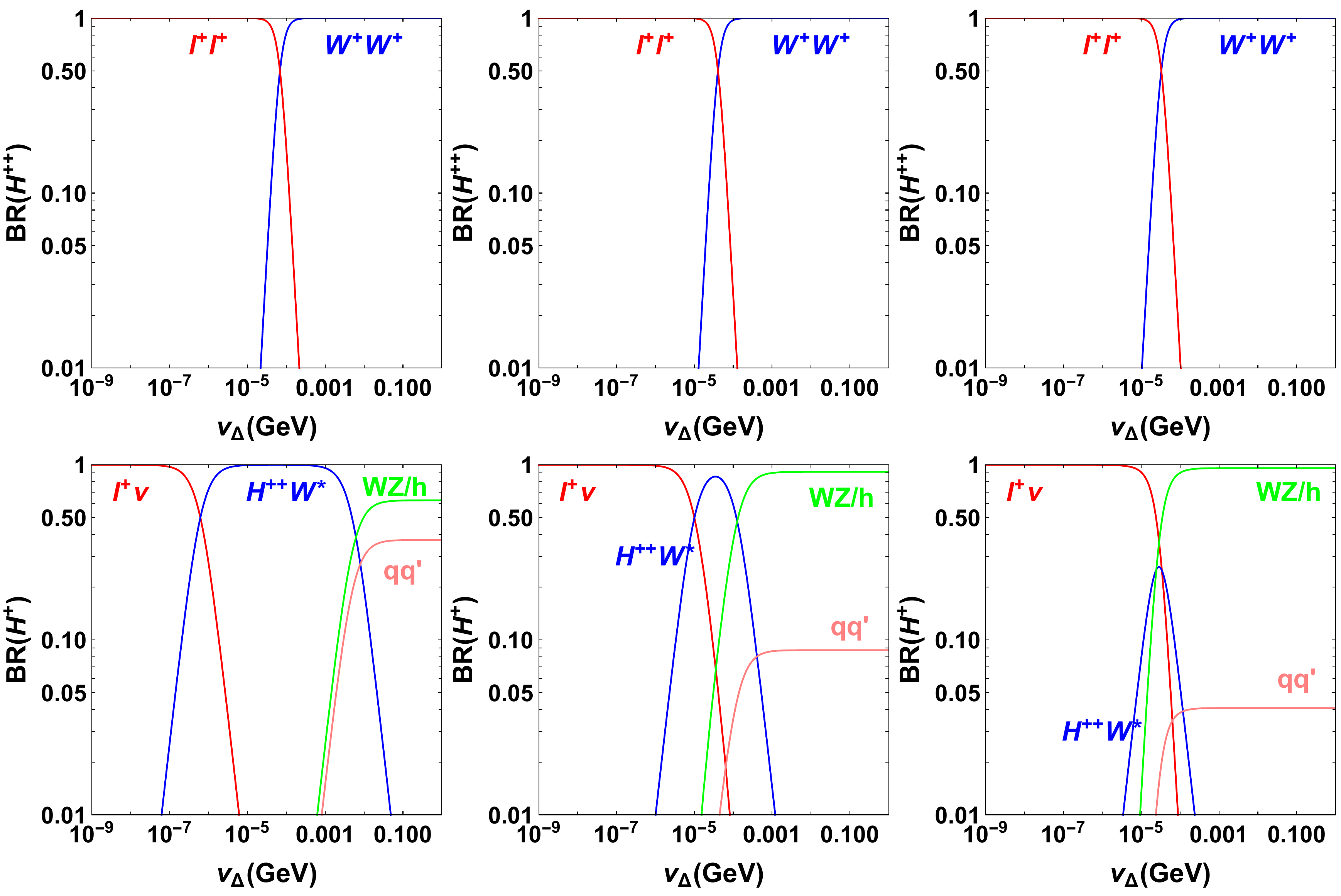}
		\caption{The branching ration of $H^{++}$ (upper panels) and $H^+$ (lower panels) for masses $M_{A^{0}} = 400$ GeV (left), 1000 GeV (middle), and 1500 GeV (right). The other relevant parameters are fixed as $\lambda_0=0.52$, $\lambda_{1,2,3}=0.1$, and $\lambda_4=0.3$.
		\label{FIG:BR1}} 	
\end{figure}

Here, we briefly discuss the decay properties of triplet scalars with the mass spectrum $M_{H^{++}}<M_{H^+}<M_{H^0}\simeq M_{A^0}$. Expressions of partial decay widths of triplet scalars can be found in Ref.~\cite{Aoki:2011pz}. In this scenario, the doubly charged Higgs $H^{\pm\pm}$ is the lightest. The possible decay channels are same-sign dilepton $H^{\pm\pm}\to \ell^\pm\ell^\pm$ and same-sign diboson $H^{\pm\pm}\to W^\pm W^\pm$.  The branching ratios are plotted in Fig.~\ref{FIG:BR1} for three benchmark cases with $M_{A^0}=$400, 1000, and 1500 GeV. The  decay widths of dilepton $H^{\pm\pm}\to \ell^\pm\ell^\pm$ channel is proportional to $1/v_{\Delta}^2$, while that of diboson $H^{\pm\pm}\to W^\pm W^\pm$ is proportional to $v_\Delta^2$. Therefore, we have BR($H^{\pm\pm}\to \ell^\pm\ell^\pm)\simeq 1$ for $v_\Delta\lesssim 10^{-5}$ GeV,  BR($H^\pm\to H^{\pm\pm}W^*)\simeq1$ for $v_\Delta\gtrsim10^{-3}$ GeV. Increasing the mass of $H^{\pm\pm}$ do not have a large impact on the results of BR($H^{\pm\pm}$). As for the singly charged Higgs $H^\pm$, possible decay channels are leptonic $H^\pm \to \ell^\pm \nu$, bosonic $H^\pm \to W^\pm Z/W^\pm h$, quarks $H^\pm \to tb/cs$, and cascade $H^\pm\to H^{\pm\pm}W^*$. Here, we focus on the same-sign tetralepton signature related channel, i.e., the cascade decay $H^\pm\to H^{\pm\pm}W^*$. This channel is the dominant one in the range of $10^{-6}~\text{GeV}\lesssim v_\Delta\lesssim 10^{-3}$ GeV when $M_{A^0}=400$ GeV. As the mass of triplet scalars increase to about 1000 GeV, the dominant range of this channel shrinks to $v_\Delta\sim5\times10^{-5}$ GeV, and the corresponding branching ratio never reaches one. Meanwhile, this channel can not become the dominant one  when $M_{A^0}=1500$ GeV. This is because of as the increase of triplet scalar masses, the phase space of cascade decay is suppressed.
It has been shown that the dominant range of cascade decays $H^0\to H^\pm W^*$ and $A^0\to H^\pm W^*$ are similar to the channel $H^\pm\to H^{\pm\pm}W^*$ \cite{Han:2015hba}. 
 
\section{Same-Sign Tetalepton Signature}\label{SEC3}
\begin{figure} 
	\centering
	\includegraphics[width=0.5\textwidth]{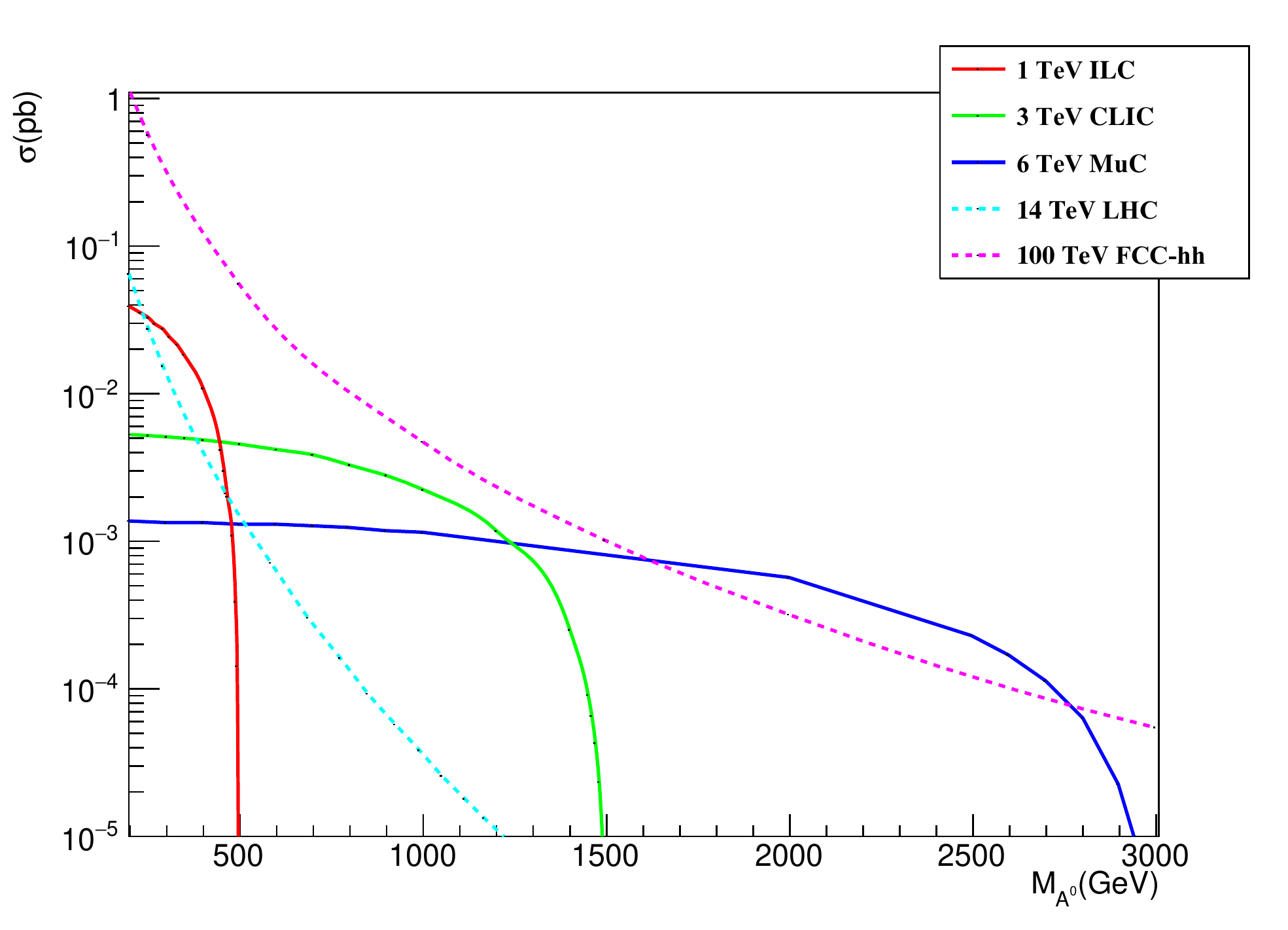}
	\caption{Production cross section of $H^0A^0$ at various colliders. The solid red, green, and blue lines are the results at 1 TeV ILC, 3 TeV CLIC, and 6 TeV MuC, respectively. The dashed cyan and pink lines are the results at 14 TeV LHC and 100 TeV FCC-hh.
		\label{FIG:XS}} 	
\end{figure}

In this section, we explore the same-sign tetralepton signature resulting from the neutral Higgs decay. First, let's consider the production cross section of $H^0A^0$. The results are shown in Fig.~\ref{FIG:XS}, where the cross section $\sigma(H^0A^0)$ at 14 TeV LHC and 100 TeV FCC-hh are also illustrated for comparison. All the results are computed by using Madgraph5\_aMC@NLO \cite{Alwall:2014hca}. For lepton colliders, the neutral Higgs pair $H^0A^0$ can be produced when $M_{A^0}<\sqrt{s}/2$. At the 1 TeV ILC, the cross section $\sigma(H^0A^0)$ is larger than at 14 TeV LHC in the range of $300~\text{GeV}\lesssim M_{A^0}\lesssim 500$ GeV. For $500~\text{GeV}\lesssim M_{A^0}\lesssim 1300$ GeV, the 3 TeV CLIC generates the largest cross section among lepton colliders. Notably, $\sigma(H^0A^0)$ at 3 TeV CLIC can be two orders of magnitudes larger than at LHC for $M_{A^0}\sim 1000$ GeV. When $M_{A^0}\gtrsim 1300$ GeV, the 6 TeV MuC becomes one of the best options. Especially, in the range of $1700~\text{GeV}\lesssim M_{A^0}\lesssim2700$ GeV, $\sigma(H^0A^0)$ at 6 TeV MuC is even larger than at 100 TeV FCC-hh.

\begin{figure} 
	\centering
	\includegraphics[width=0.33\textwidth]{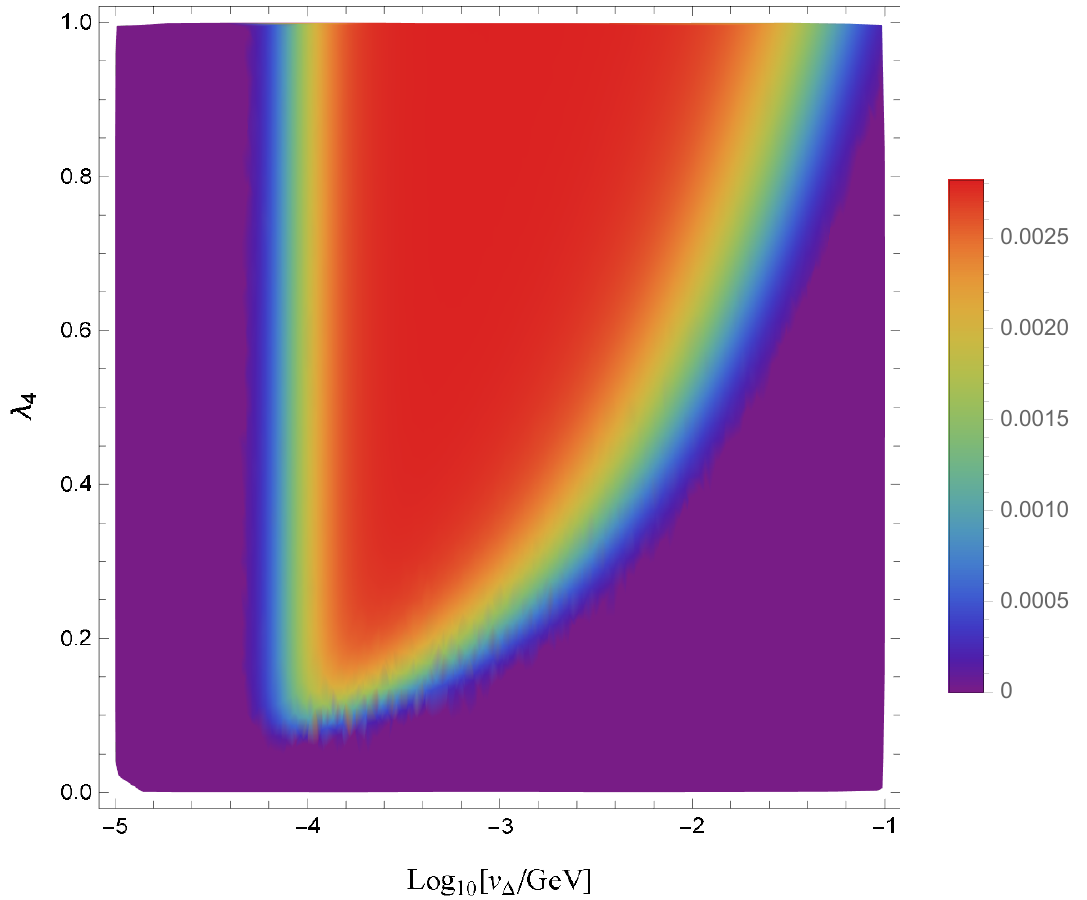}
	\includegraphics[width=0.32\textwidth]{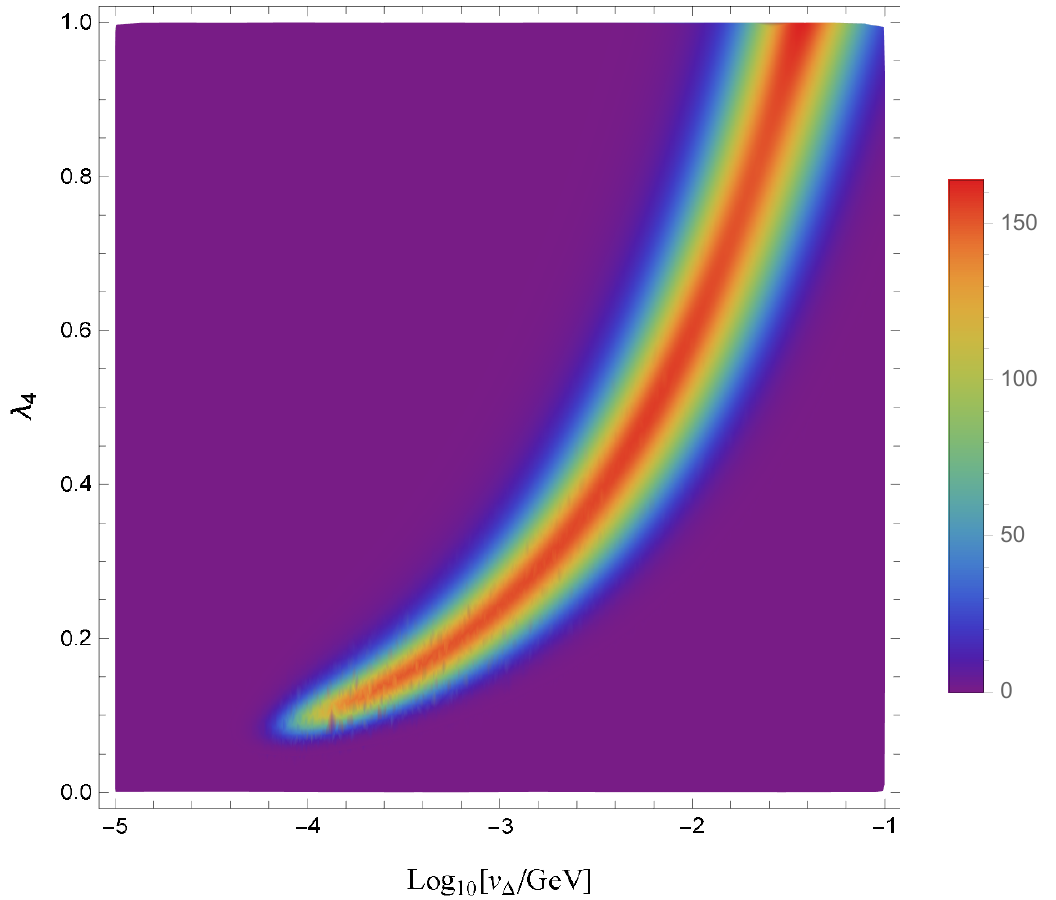}
	\caption{Left panel: Product of branching ratios $\operatorname{BR}\left(H^{0}/A^0 \rightarrow H^{\pm} W^{-*}\right)^2 \times \operatorname{BR}\left(H^{\pm} \rightarrow H^{\pm \pm} W^{-*}\right)^{2}  \times \operatorname{BR}\left(H^{\pm \pm} \rightarrow W^{\pm} W^{\pm}\right)^{2} \times \operatorname{BR}\left(W^{\pm} \rightarrow \ell \nu\right)^{4}$ for the process ~$e^+e^- \rightarrow H^0 A^0$~ with mass of ~$A^0$~ being fixed as ~$M_{A^0}=400~ \mathrm{GeV}$.
	 Right panel: Event number of the same-sign tetralepton signature $4\ell^{\pm} +X$ for the mass $M_{A^{0}}= 400 ~\mathrm{GeV}$ from ~$e^+e^- \to H^0 A^0$~ and subsequent decays at the  $\sqrt{s}=1~ \mathrm{TeV}$ ILC with  $\mathcal{L}=8~ \mathrm{ab}^{-1}$.
		\label{FIG:BP1}} 	
\end{figure}

At the 1 TeV ILC with $M_{A^0}=400$ GeV, this signal is generated via the tetraboson process
\begin{eqnarray}\label{Eqn:PR1}
	e^+e^- \to H^0A^0 &\to& H^\pm W^{ *}H^\pm W^{ *} \to  H^{\pm\pm} W^{ *} H^{\pm\pm} W^{ *}+ W^{ *} W^{ *}\to 4 W^\pm  + X,
\end{eqnarray}
with the leptonic decay $W^\pm\to \ell^\pm \nu ~(\ell=e,\mu)$. Note that the dilepton decay $H^{\pm\pm}\to \ell^\pm\ell^\pm$ has already been excluded by direct search at LHC. Since the typical mass splitting between triplet scalars for the same-sign tetralepton signature is at the order of $\mathcal{O}(\text{GeV})$, the final states from off-shell $W$ decay are hard to be detected.  Such signature occurs due to the interference effect between $H^0$ and $A^0$, which is sizable when $\delta M=M_{H^0}-M_{A^0}\sim \Gamma_{H^0/A^0}$. The cross section for this signal is calculated as \cite{Chun:2012zu}
\begin{eqnarray} \label{Eqn:CS1}
\sigma_W(4\ell^\pm+X) &=&
\sigma\left(e^+ e^- \rightarrow H^{0} A^{0}\right) \times \left(\frac{2+x^2}{1+x^2}\frac{x^2}{1+x^2}\right) \times \text{BR}\left(H^{0} / A^{0} \rightarrow H^{\pm} W^{*}\right)^{2} \\ \nonumber
&&\times  \text{BR}\left(H^{\pm} \rightarrow H^{\pm \pm} W^{*}\right)^{2} \times \text{BR}\left(H^{\pm \pm} \rightarrow  W^{\pm} W^{\pm}\right)^{2} \times \text{BR}(W^\pm\to \ell^\pm\nu)^4,
\end{eqnarray}
where $x=\delta M/\Gamma_{H^0/A^0}$. The initial cross section $\sigma(e^+e^-\to H^0A^0)$ is about 10 fb at the 1 TeV ILC with $M_{A^0}=400$ GeV. In the left panel of Fig.~\ref{FIG:BP1}, we show the product of BRs in the above process. As shown in Fig.~\ref{FIG:BR1}, BR$(H^{\pm\pm}\to W^\pm W^\pm)$ is quickly suppressed for $v_\Delta <10^{-4}$ GeV, which corresponds to the left boundary. While the right one is determined by the cascade decay branching ratios as BR$(H^\pm\to H^{\pm\pm}W^*)$. In this way, a larger $\lambda_4$ leads to a larger mass splitting, hence a wider range of $v_\Delta$  \cite{Han:2015sca}.

In the right panel of Fig.~\ref{FIG:BP1}, we show the expected event number for the same-sign tetralepton signature at the 1 TeV ILC with an integrated luminosity of $\mathcal{L}=8~\text{ab}^{-1}$. A detector level simulation with Delphes \cite{deFavereau:2013fsa} is also performed, where we only require $p_T(\ell^\pm)>10$ GeV and $|\eta(\ell^\pm)|<2.5$. The total cut efficiency we applied is $c_{eff}=0.6$ for $M_{A^0}=400$ GeV. The promising region in the $\lambda_4-v_\Delta$ plane fills a narrow band, where the maximum event number can reach about 160.  Such a narrow band is formed mainly due to the interference effect between $H^0$ and $A^0$. For fixed value of $v_\Delta$, the mass splitting  $\delta M$ is then determined. A certain value $\lambda_4^M$ resulting suitable cascade decay width, i.e., $x=\delta M/\Gamma_{H^0/A^0}\sim1$, leads to the maximum event number. If $\lambda_4>\lambda_4^M$, then $\Gamma_{H^0/A^0}$ will increase, thus $x$ will decrease, and the final event number also will decrease. Considering the fact that for a small mass splitting of triplet scalars $\Delta M \sim \lambda_4 v^2/(8 M_{A^0})$, the cascade decay dominant width $\Gamma_{H^0/A^0}\propto \Delta M^5$, and $\delta M\propto v_\Delta^2$, it is easy to derive the relation $\lambda_4\propto v_\Delta^{2/5}$ by taking $\delta M\sim \Gamma_{H^0/A^0}$.

Now, let's consider the same-sign tetralepton signature at the 3 TeV CLIC. In this scenario, we set $M_{A^0}=1000$ GeV, and the same-sign dilepton decay $H^{\pm}\to \ell^\pm \ell^\pm$ is still allowed. Therefore, in addition to the tetraboson process in Eqn.~\eqref{Eqn:PR1}, we also have the direct tetralepton channel
\begin{eqnarray}\label{Eqn:PR2}
e^+e^- \to H^0A^0 &\to& H^\pm W^{ *}H^\pm W^{ *} \to  H^{\pm\pm} W^{ *} H^{\pm\pm} W^{ *}+ W^{ *} W^{ *}\to 4 \ell^\pm  + X.
\end{eqnarray}
The corresponding cross section is then calculated as
\begin{eqnarray} \label{Eqn:CS2}
\sigma_\ell(4\ell^\pm+X) &=&
\sigma\left(e^+ e^- \rightarrow H^{0} A^{0}\right) \times \left(\frac{2+x^2}{1+x^2}\frac{x^2}{1+x^2}\right) \times \text{BR}\left(H^{0} / A^{0} \rightarrow H^{\pm} W^{*}\right)^{2} \\ \nonumber
&&\times  \text{BR}\left(H^{\pm} \rightarrow H^{\pm \pm} W^{*}\right)^{2} \times \text{BR}\left(H^{\pm \pm} \rightarrow  \ell^{\pm} \ell^{\pm}\right)^{2}.
\end{eqnarray}

\begin{figure} 
	\centering
	\includegraphics[width=0.33\textwidth]{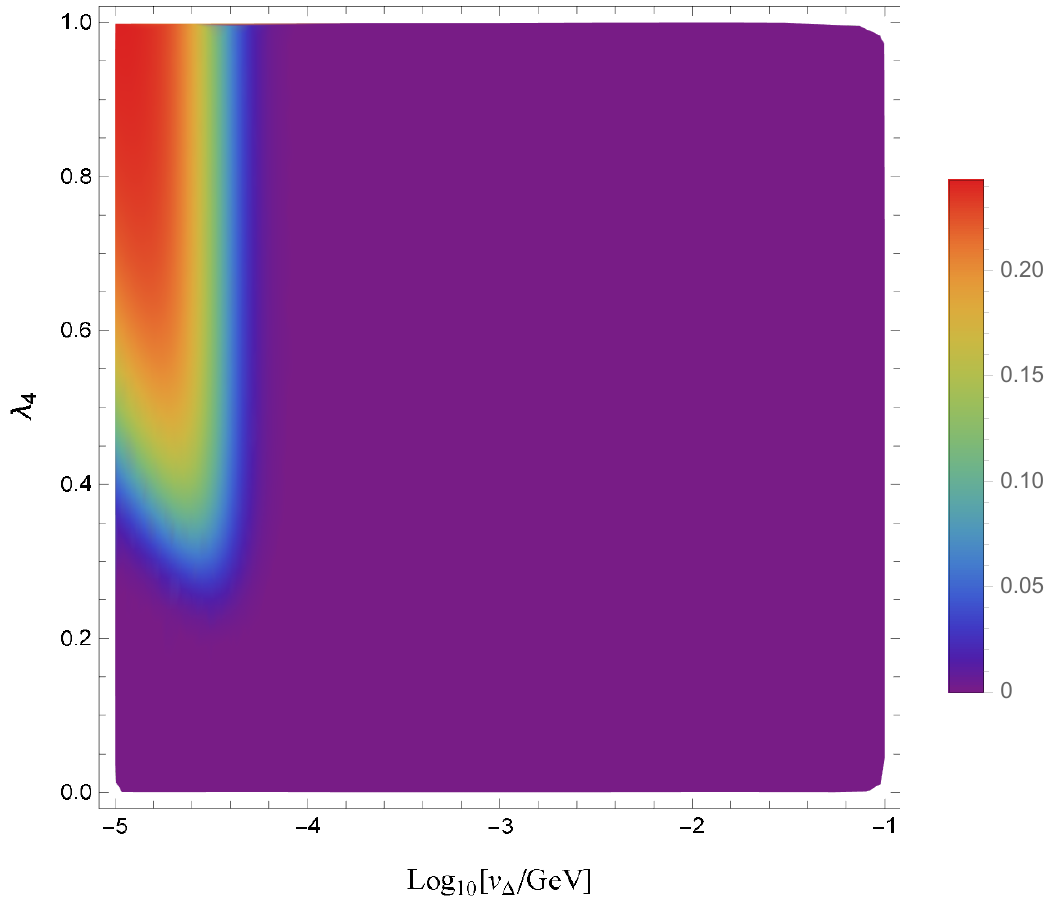}
	\includegraphics[width=0.335\textwidth]{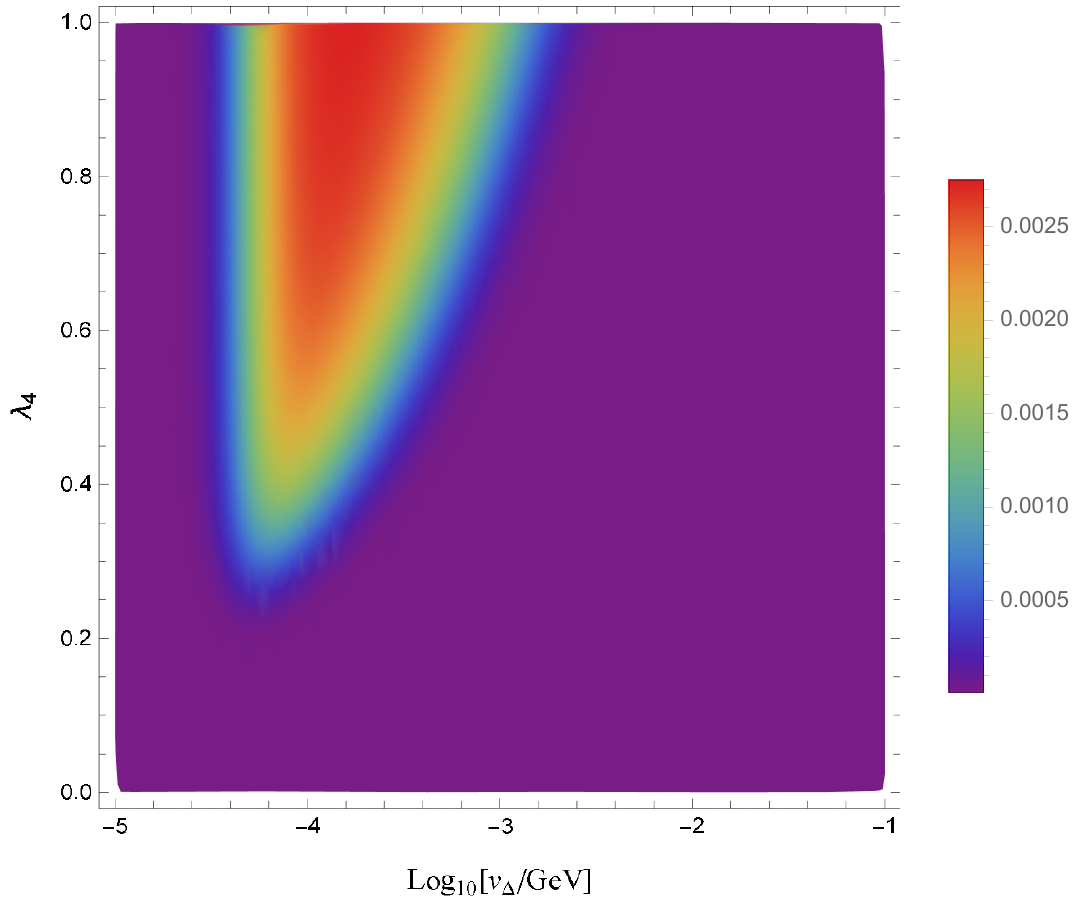}
	\includegraphics[width=0.325\textwidth]{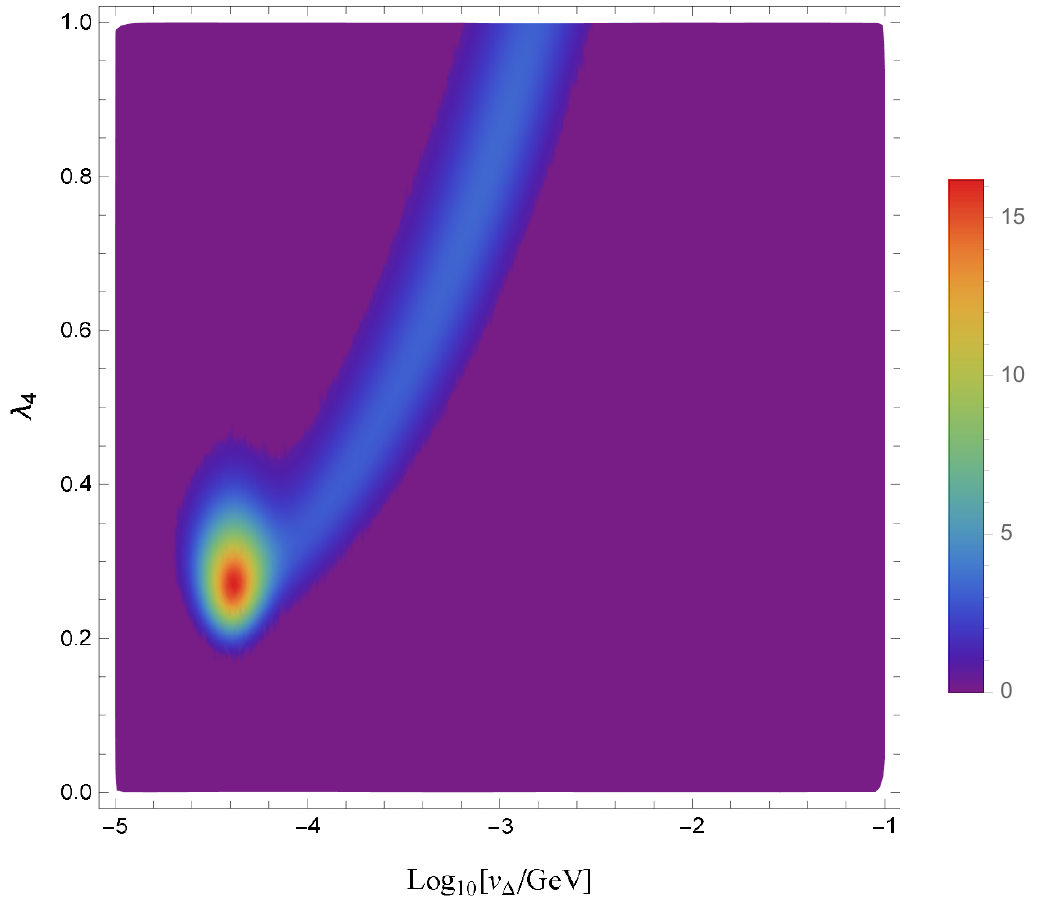}
	\caption{Left panel: Product of branching ratios $\operatorname{BR}\left(H^{0}/A^0 \rightarrow H^{\pm} W^{-*}\right)^2 \times \operatorname{BR}\left(H^{\pm} \rightarrow H^{\pm \pm} W^{-*}\right)^{2}  \times \operatorname{BR}\left(H^{\pm \pm} \rightarrow \ell^{\pm} \ell^{\pm}\right)^{2}$ for the process ~$e^+e^- \rightarrow H^0 A^0$~ with $M_{A^0}=1000 \mathrm{GeV}$.
	Middle panel: Product of branching ratios $\operatorname{BR}\left(H^{0}/A^0 \rightarrow H^{\pm} W^{-*}\right)^2 \times \operatorname{BR}\left(H^{\pm} \rightarrow H^{\pm \pm} W^{-*}\right)^{2}  \times \operatorname{BR}\left(H^{\pm \pm} \rightarrow W^{\pm} W^{\pm}\right)^{2} \times \operatorname{BR}\left(W^{\pm} \rightarrow \ell \nu\right)^{4}$.
	Right panel: Event number of the same-sign tetralepton signature $4\ell^{\pm}+X$ for the mass $M_{H^{0}} \sim M_{A^{0}}= 1000 \mathrm{GeV}$ from ~$e^+e^- \to H^0 A^0$~ and subsequent decays at the  $\sqrt{s}=3 ~\mathrm{TeV}$ CLIC with luminosity $\mathcal{L}=5~ \mathrm{ab}^{-1}$.
		\label{FIG:BP2}} 	
\end{figure}

In the left panel of Fig.~\ref{FIG:BP2}, we show the product of BRs in the direct tetralepton decay process. As shown in Fig.~\ref{FIG:BR1}, the cascade decays are suppressed for $v_\Delta\lesssim 10^{-5}$ GeV with $M_{A^0}=1000$ GeV, so we do not show the region $v_\Delta<10^{-5}$ GeV. The right boundary corresponds to the area where BR($H^{\pm\pm}\to\ell^\pm\ell^\pm$) is suppressed. For $\lambda_4>0.5$, there are large parameter space where the product of BRs reaches the maximum, i.e., 0.25. In the middle panel of Fig.~\ref{FIG:BP2}, the product of BRs in the diboson process is also shown. Comparing with the region of $M_{A^0}=400$ GeV in Fig.~\ref{FIG:BP1}, the region of $M_{A^0}=1000$ GeV is much smaller. For instance, when product of BRs is larger than 0.002, one needs $\lambda_4\gtrsim 0.5$ and $10^{-4}~\text{GeV}\lesssim v_\Delta\lesssim10^{-3}$ GeV. This is because  for heavier scalar triplet, the branching ratios of cascade decays are suppressed. 

In the right panel of Fig.~\ref{FIG:BP2}, we show the expected event number for the same-sign tetralepton signature at the 3 TeV CLIC with an integrated luminosity of $\mathcal{L}=5~\text{ab}^{-1}$. Here, the expected event number is the sum of both diboson decay process in Eqn.~\eqref{Eqn:CS1} and the dilepton decay process in Eqn.~\ref{Eqn:CS2}. In a small area around $v_\Delta\sim4\times10^{-4}$ GeV and $\lambda_4\sim0.26$, we have the maximum number $\sim 16$, where the dominant contribution is from $H^{\pm\pm}\to \ell^\pm\ell^\pm$.  Meanwhile, the $H^{\pm\pm}\to W^\pm W^\pm$ dominant tail region with $10^{-4}~\text{GeV}\lesssim v_\Delta\lesssim 10^{-3}$ GeV only predicts a total event number less than three, thus this long tail region is not promising.  

\begin{figure} 
	\centering
	\includegraphics[width=0.33\textwidth]{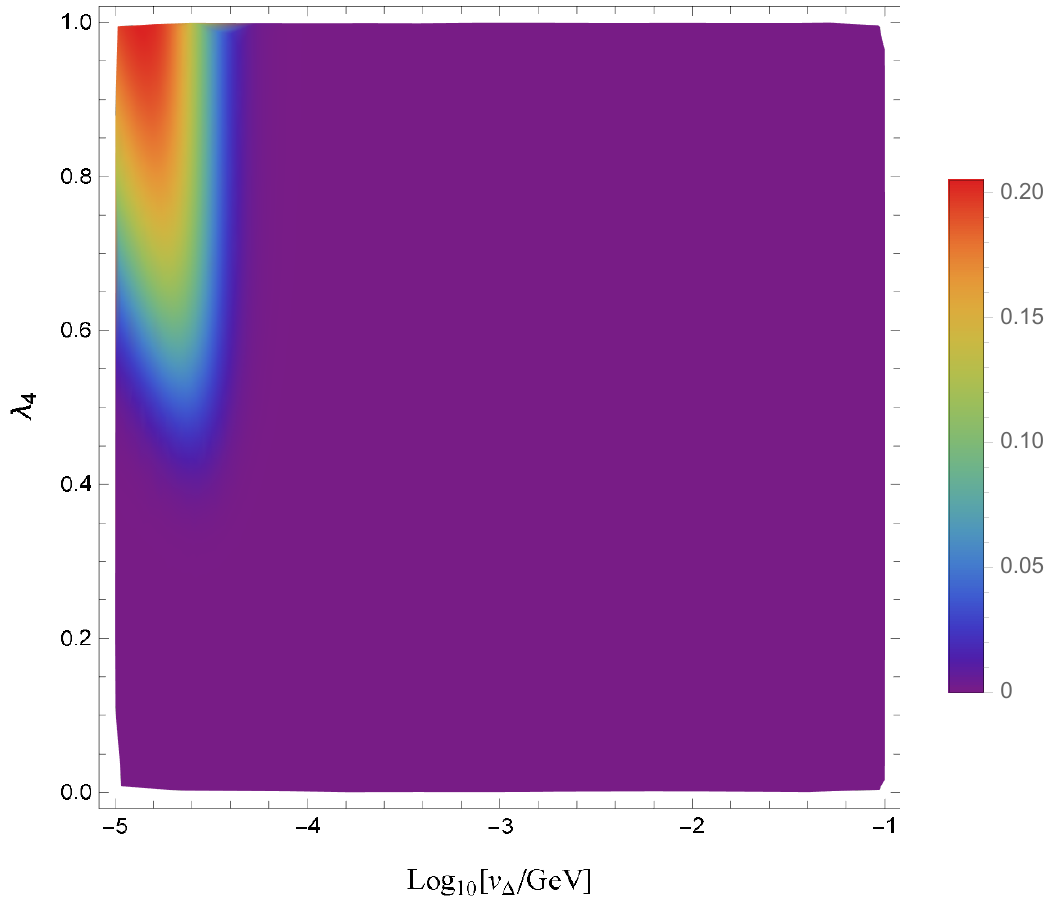}
	\includegraphics[width=0.335\textwidth]{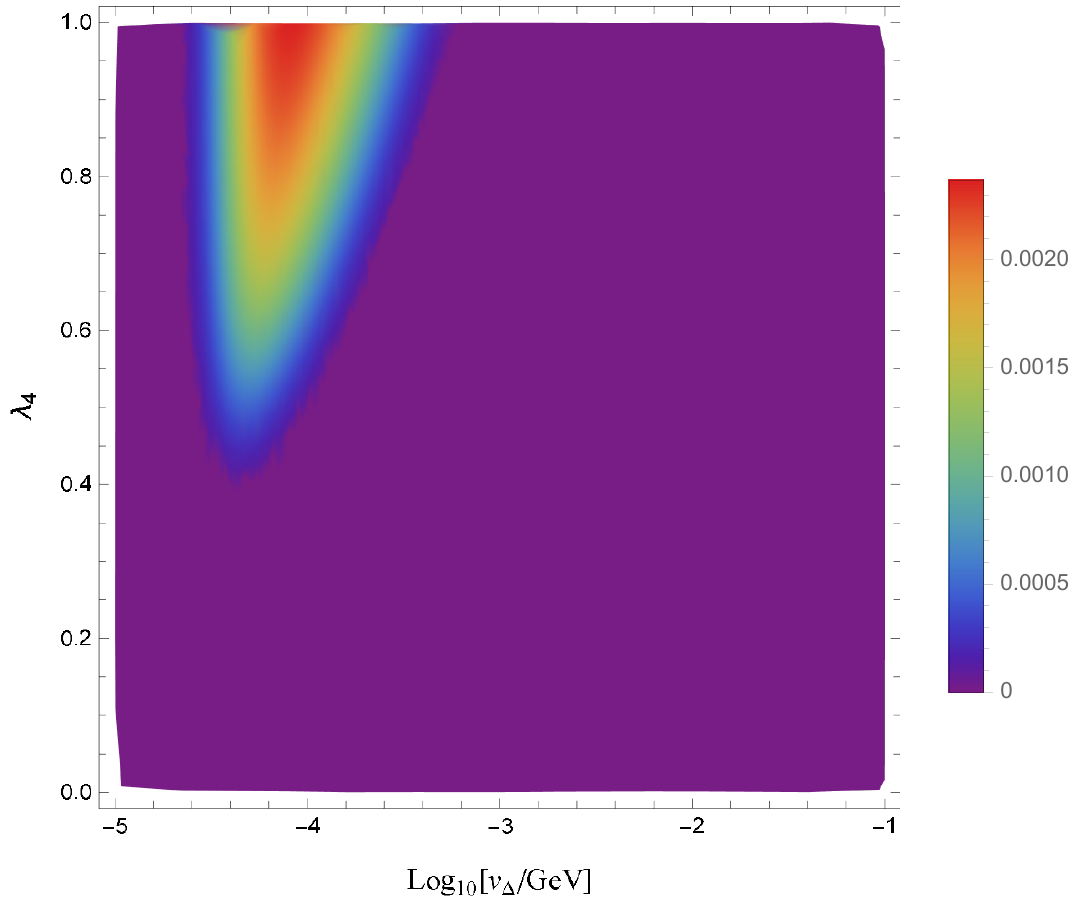}
	\includegraphics[width=0.325\textwidth]{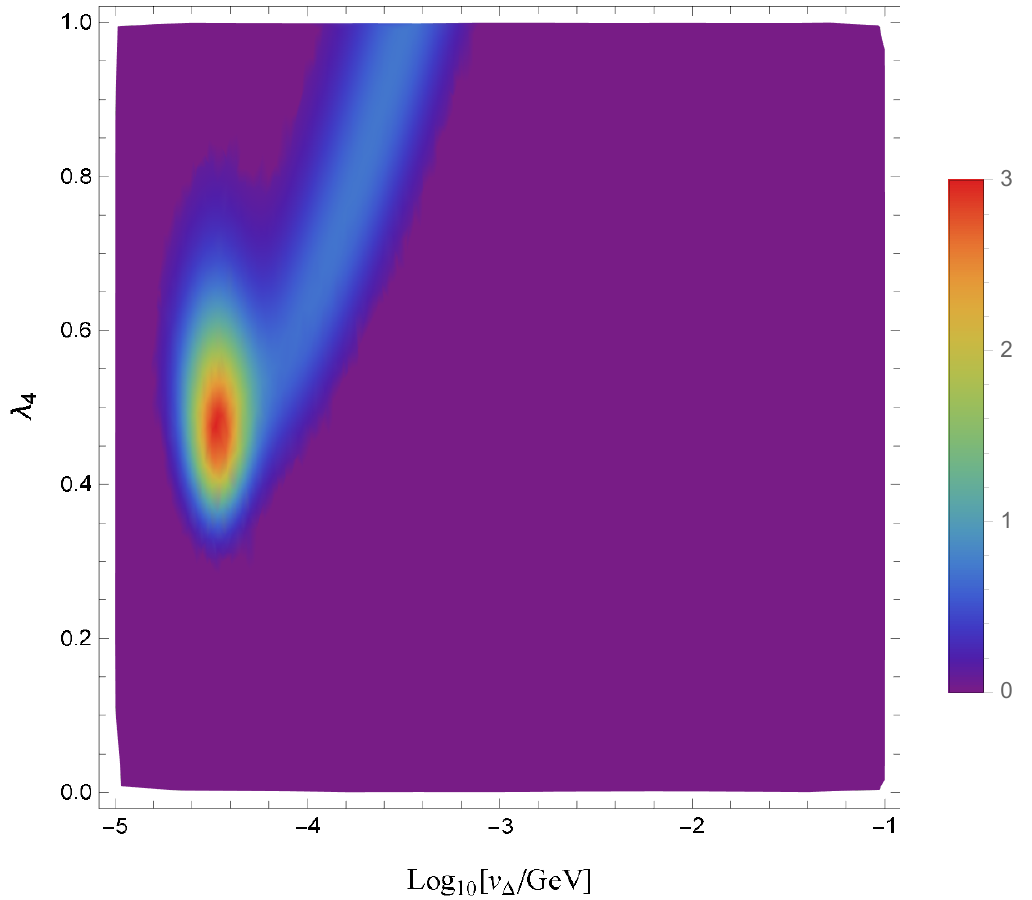}
	\caption{Same as Fig.~\ref{FIG:BP2}, but for  $M_{H^{0}} \sim M_{A^{0}}= 1500 \mathrm{GeV}$ from the process $\mu^+\mu^-\to H^0A^0$ at the  $\sqrt{s}=6 ~\mathrm{TeV}$ MuC with luminosity $\mathcal{L}=10~ \mathrm{ab}^{-1}$.
		\label{FIG:BP3}} 	
\end{figure}

At last, we consider the same-sign tetralepton signature at the 6 TeV MuC. The corresponding production processes at muon collider are 
\begin{eqnarray}
\mu^+\mu^- \to H^0A^0 &\to& H^\pm W^{ *}H^\pm W^{ *} \to  H^{\pm\pm} W^{ *} H^{\pm\pm} W^{ *}+ W^{ *} W^{ *}\to 4 W^\pm(\to \ell^\pm\nu)  + X,\\
\mu^+\mu^- \to H^0A^0 &\to& H^\pm W^{ *}H^\pm W^{ *} \to  H^{\pm\pm} W^{ *} H^{\pm\pm} W^{ *}+ W^{ *} W^{ *}\to 4 \ell^\pm  + X.
\end{eqnarray}
The production cross section is then obtained by simply replace $\sigma(e^+e^-\to H^0A^0)$ in Eqn.~\ref{Eqn:CS1} and Eqn.~\ref{Eqn:CS2} with $\sigma(\mu^+\mu^-\to H^0A^0)$. In the left and middle panel of Fig.~\ref{FIG:BP3}, we show the product of BRs in the direct tetralepton and tetraboson decay process with $M_{A^0}=1500$ GeV. To realize a relatively large value of BRs, $\lambda_4$ has to be larger than $0.8$. However, such large $\lambda_4$ leads to too large mass splitting of triplet scalars that the interference factor $x$ is suppressed. In the right panel of Fig.~\ref{FIG:BP3}, we show the total event number for the same-sign tetralepton signature at the 6 TeV MuC with an integrated luminosity of $\mathcal{L}=10~\text{ab}^{-1}$. It is obvious that the event number is always smaller than three. Therefore, the same-sign tetralepton signature is not promising at the MuC for $M_{A^0}=1500$ GeV. 

\section{Conclusion}\label{SEC4}

In this paper, we study the novel same-sign tetra-lepton signature in type-II seesaw at the future lepton colliders~(including 1 TeV ILC, 3 TeV CLIC, and 6 TeV MuC). The signature arises from the mixing of associated production of Higgs fields $H^0 A^0$ followed by the cascade decays $H^{0} / A^{0} \rightarrow H^{\pm} W^{*}$, $H^{\pm} \rightarrow H^{\pm \pm} W^{*}$, and $H^{\pm\pm}\to \ell^\pm\ell^\pm/W^\pm W^\pm$ with $W^\pm\to \ell^\pm\nu$. There are two important parameters $\lambda_4$ and $v_\Delta$ closely related to this signature, where $\lambda_4$ controls the mass splitting of triplet scalars and $v_\Delta$ determines the decay mode of $H^{\pm\pm}$.

We first consider a low mass benchmark scenario with $M_{A^0}=400$ GeV at 1 TeV ILC. In this scenario, $H^{\pm\pm}\to W^\pm W^\pm$ is the only viable decay mode. The production cross section of the process $e^+ e^- \to H^0 A^0$ varies around 10 fb. The promising region corresponds to a narrow band in the range of $10^{-4}~\text{GeV}\lesssim v_\Delta \lesssim10^{-2}$ GeV.  With an integrated luminosity of $\mathcal{L}=8~ \mathrm{ab}^{-1}$, we find that a neutral Higgs of mass around ~$400 ~\mathrm{GeV}$~ can lead to around $150$ events at ILC. For heavier triplet scalars, we then consider $M_{A^0}=1000$ GeV at 3 TeV CLIC, where the cross section  $\sigma(e^+ e^- \to H^0 A^0)$ is about 2 fb. Although this value is about two orders of magnitudes larger than at 14 TeV LHC, the cascade decay branching ratios are suppressed for small $\lambda_4$. This leads to a miss match between cascade decays and the interference effect. A maximum event number $\sim 16$ can be obtained around  $v_\Delta\sim4\times10^{-4}$ GeV and $\lambda_4\sim0.26$ with an integrated luminosity of $\mathcal{L}=5~ \mathrm{ab}^{-1}$ at CLIC. In this high mass scenario, the $H^{\pm\pm}\to \ell^\pm\ell^\pm$ decay mode is the dominant contribution to the same-sign tetralepton signature. If the triplet scalars are even heavier than 1 TeV, e.g., $M_{A^0}=1500$ GeV, the cascade decays will be heavily suppressed. With an integrated luminosity of $\mathcal{L}=10~ \mathrm{ab}^{-1}$ at 6 TeV MuC, there are at best have 3 signal events. Therefore, this signature is not promising at MuC.

\section*{Acknowledgments}

This work is supported by the National Natural Science Foundation of China under Grant No. 11605074 and 11805081, Natural Science Foundation of Shandong Province under Grant No. ZR2019QA021 and  ZR2018MA047, Joint Large-Scale Scientific Facility Funds of the NSFC and CAS under Contracts Nos. U1732263 and U2032115.


\begin{thebibliography}{000}

\bibitem{Fukuda:1998mi}
Y.~Fukuda \textit{et al.} [Super-Kamiokande],
Phys. Rev. Lett. \textbf{81}, 1562-1567 (1998)
[arXiv:hep-ex/9807003 [hep-ex]].

\bibitem{Ahmad:2002jz}
Q.~R.~Ahmad \textit{et al.} [SNO],
Phys. Rev. Lett. \textbf{89}, 011301 (2002)
[arXiv:nucl-ex/0204008 [nucl-ex]].

\bibitem{An:2012eh}
F.~P.~An \textit{et al.} [Daya Bay],
Phys. Rev. Lett. \textbf{108}, 171803 (2012)
[arXiv:1203.1669 [hep-ex]].

\bibitem{Weinberg:1979sa}
S.~Weinberg,
Phys. Rev. Lett. \textbf{43}, 1566-1570 (1979)

\bibitem{Ma:1998dn}
E.~Ma,
Phys. Rev. Lett. \textbf{81}, 1171-1174 (1998)
[arXiv:hep-ph/9805219 [hep-ph]].

\bibitem{Minkowski:1977sc}
P.~Minkowski,
Phys. Lett. B \textbf{67}, 421-428 (1977)

\bibitem{Mohapatra:1979ia}
R.~N.~Mohapatra and G.~Senjanovic,
Phys. Rev. Lett. \textbf{44}, 912 (1980)

\bibitem{Magg:1980ut}
M.~Magg and C.~Wetterich,
Phys. Lett. B \textbf{94}, 61-64 (1980)

\bibitem{Cheng:1980qt}
T.~P.~Cheng and L.~F.~Li,
Phys. Rev. D \textbf{22}, 2860 (1980)

\bibitem{Lazarides:1980nt}
G.~Lazarides, Q.~Shafi and C.~Wetterich,
Nucl. Phys. B \textbf{181}, 287-300 (1981)

\bibitem{Mohapatra:1980yp}
R.~N.~Mohapatra and G.~Senjanovic,
Phys. Rev. D \textbf{23}, 165 (1981)

\bibitem{Foot:1988aq}
R.~Foot, H.~Lew, X.~G.~He and G.~C.~Joshi,
Z. Phys. C \textbf{44}, 441 (1989)

\bibitem{delAguila:2008cj}
F.~del Aguila and J.~A.~Aguilar-Saavedra,
Nucl. Phys. B \textbf{813}, 22-90 (2009)
[arXiv:0808.2468 [hep-ph]].

\bibitem{Deppisch:2015qwa}
F.~F.~Deppisch, P.~S.~Bhupal Dev and A.~Pilaftsis,
New J. Phys. \textbf{17}, no.7, 075019 (2015)
[arXiv:1502.06541 [hep-ph]].

\bibitem{Cai:2017mow}
Y.~Cai, T.~Han, T.~Li and R.~Ruiz,
Front. in Phys. \textbf{6}, 40 (2018)
[arXiv:1711.02180 [hep-ph]].

\bibitem{Boucenna:2014zba}
S.~M.~Boucenna, S.~Morisi and J.~W.~F.~Valle,
Adv. High Energy Phys. \textbf{2014}, 831598 (2014)
[arXiv:1404.3751 [hep-ph]].

\bibitem{Cai:2017jrq}
Y.~Cai, J.~Herrero-Garc\'\i{}a, M.~A.~Schmidt, A.~Vicente and R.~R.~Volkas,
Front. in Phys. \textbf{5}, 63 (2017)
[arXiv:1706.08524 [hep-ph]].

\bibitem{Arhrib:2011uy}
A.~Arhrib, R.~Benbrik, M.~Chabab, G.~Moultaka, M.~C.~Peyranere, L.~Rahili and J.~Ramadan,
Phys. Rev. D \textbf{84}, 095005 (2011)
[arXiv:1105.1925 [hep-ph]].

\bibitem{Perez:2008ha}
P.~Fileviez Perez, T.~Han, G.~y.~Huang, T.~Li and K.~Wang,
Phys. Rev. D \textbf{78}, 015018 (2008)
[arXiv:0805.3536 [hep-ph]].

\bibitem{Melfo:2011nx}
A.~Melfo, M.~Nemevsek, F.~Nesti, G.~Senjanovic and Y.~Zhang,
Phys. Rev. D \textbf{85}, 055018 (2012)
[arXiv:1108.4416 [hep-ph]].

\bibitem{Aoki:2011pz}
M.~Aoki, S.~Kanemura and K.~Yagyu,
Phys. Rev. D \textbf{85}, 055007 (2012)
[arXiv:1110.4625 [hep-ph]].

\bibitem{Han:2015hba}
Z.~L.~Han, R.~Ding and Y.~Liao,
Phys. Rev. D \textbf{91}, 093006 (2015)
[arXiv:1502.05242 [hep-ph]].

\bibitem{Han:2015sca}
Z.~L.~Han, R.~Ding and Y.~Liao,
Phys. Rev. D \textbf{92}, no.3, 033014 (2015)
[arXiv:1506.08996 [hep-ph]].

\bibitem{Akeroyd:2011zza}
A.~G.~Akeroyd and H.~Sugiyama,
Phys. Rev. D \textbf{84}, 035010 (2011)
[arXiv:1105.2209 [hep-ph]].

\bibitem{Akeroyd:2012nd}
A.~G.~Akeroyd, S.~Moretti and H.~Sugiyama,
Phys. Rev. D \textbf{85}, 055026 (2012)
[arXiv:1201.5047 [hep-ph]].

\bibitem{Chun:2013vma}
E.~J.~Chun and P.~Sharma,
Phys. Lett. B \textbf{728}, 256-261 (2014)
[arXiv:1309.6888 [hep-ph]].

\bibitem{Babu:2016rcr}
K.~S.~Babu and S.~Jana,
Phys. Rev. D \textbf{95}, no.5, 055020 (2017)
[arXiv:1612.09224 [hep-ph]].

\bibitem{Li:2018jns}
T.~Li,
JHEP \textbf{09}, 079 (2018)
[arXiv:1802.00945 [hep-ph]].

\bibitem{Primulando:2019evb}
R.~Primulando, J.~Julio and P.~Uttayarat,
JHEP \textbf{08}, 024 (2019)
[arXiv:1903.02493 [hep-ph]].

\bibitem{Du:2018eaw}
Y.~Du, A.~Dunbrack, M.~J.~Ramsey-Musolf and J.~H.~Yu,
JHEP \textbf{01}, 101 (2019)
[arXiv:1810.09450 [hep-ph]].

\bibitem{deMelo:2019asm}
T.~B.~de Melo, F.~S.~Queiroz and Y.~Villamizar,
Int. J. Mod. Phys. A \textbf{34}, no.27, 1950157 (2019)
[arXiv:1909.07429 [hep-ph]].

\bibitem{Padhan:2019jlc}
R.~Padhan, D.~Das, M.~Mitra and A.~Kumar Nayak,
Phys. Rev. D \textbf{101}, no.7, 075050 (2020)
[arXiv:1909.10495 [hep-ph]].

\bibitem{Fuks:2019clu}
B.~Fuks, M.~Nemev\v{s}ek and R.~Ruiz,
Phys. Rev. D \textbf{101}, no.7, 075022 (2020)
[arXiv:1912.08975 [hep-ph]].

\bibitem{Blunier:2016peh}
S.~Blunier, G.~Cottin, M.~A.~D\'\i{}az and B.~Koch,
Phys. Rev. D \textbf{95}, no.7, 075038 (2017)
[arXiv:1611.07896 [hep-ph]].

\bibitem{Agrawal:2018pci}
P.~Agrawal, M.~Mitra, S.~Niyogi, S.~Shil and M.~Spannowsky,
Phys. Rev. D \textbf{98}, no.1, 015024 (2018)
[arXiv:1803.00677 [hep-ph]].

\bibitem{Dev:2019hev}
P.~S.~B.~Dev, S.~Khan, M.~Mitra and S.~K.~Rai,
Phys. Rev. D \textbf{99}, no.11, 115015 (2019)
[arXiv:1903.01431 [hep-ph]].

\bibitem{Yang:2021skb}
X.~H.~Yang and Z.~J.~Yang,
[arXiv:2103.11412 [hep-ph]].

\bibitem{Aaboud:2017qph}
M.~Aaboud \textit{et al.} [ATLAS],
Eur. Phys. J. C \textbf{78}, no.3, 199 (2018)
[arXiv:1710.09748 [hep-ex]].

\bibitem{Akeroyd:2007zv}
A.~G.~Akeroyd, M.~Aoki and H.~Sugiyama,
Phys. Rev. D \textbf{77}, 075010 (2008)
[arXiv:0712.4019 [hep-ph]].

\bibitem{Aaboud:2018qcu}
M.~Aaboud \textit{et al.} [ATLAS],
Eur. Phys. J. C \textbf{79}, no.1, 58 (2019)
[arXiv:1808.01899 [hep-ex]].

\bibitem{Aad:2021lzu}
G.~Aad \textit{et al.} [ATLAS],
[arXiv:2101.11961 [hep-ex]].

\bibitem{Chun:2012zu}
E.~J.~Chun and P.~Sharma,
JHEP \textbf{08}, 162 (2012)
[arXiv:1206.6278 [hep-ph]].

\bibitem{Chun:2019hce}
E.~J.~Chun, S.~Khan, S.~Mandal, M.~Mitra and S.~Shil,
Phys. Rev. D \textbf{101}, no.7, 075008 (2020)
[arXiv:1911.00971 [hep-ph]].


\bibitem{CEPCStudyGroup:2018ghi}
J.~B.~Guimar\~aes da Costa \textit{et al.} [CEPC Study Group],
[arXiv:1811.10545 [hep-ex]].

\bibitem{Barklow:2015tja}
T.~Barklow, J.~Brau, K.~Fujii, J.~Gao, J.~List, N.~Walker and K.~Yokoya,
[arXiv:1506.07830 [hep-ex]].

\bibitem{Fujii:2017vwa}
K.~Fujii, C.~Grojean, M.~E.~Peskin, T.~Barklow, Y.~Gao, S.~Kanemura, H.~Kim, J.~List, M.~Nojiri and M.~Perelstein, \textit{et al.}
[arXiv:1710.07621 [hep-ex]].

\bibitem{Linssen:2012hp}
L.~Linssen, A.~Miyamoto, M.~Stanitzki and H.~Weerts,
[arXiv:1202.5940 [physics.ins-det]].

\bibitem{Robson:2018zje}
A.~Robson and P.~Roloff,
[arXiv:1812.01644 [hep-ex]].

\bibitem{Delahaye:2019omf}
J.~P.~Delahaye, M.~Diemoz, K.~Long, B.~Mansouli\'e, N.~Pastrone, L.~Rivkin, D.~Schulte, A.~Skrinsky and A.~Wulzer,
[arXiv:1901.06150 [physics.acc-ph]].

\bibitem{Long:2020wfp}
K.~Long, D.~Lucchesi, M.~Palmer, N.~Pastrone, D.~Schulte and V.~Shiltsev,
[arXiv:2007.15684 [physics.acc-ph]].

\bibitem{Alwall:2014hca}
J.~Alwall, R.~Frederix, S.~Frixione, V.~Hirschi, F.~Maltoni, O.~Mattelaer, H.~S.~Shao, T.~Stelzer, P.~Torrielli and M.~Zaro,
JHEP \textbf{07}, 079 (2014)
[arXiv:1405.0301 [hep-ph]].

\bibitem{deFavereau:2013fsa}
J.~de Favereau \textit{et al.} [DELPHES 3],
JHEP \textbf{02}, 057 (2014)
[arXiv:1307.6346 [hep-ex]].

\end{thebibliography}
\end{document}